\documentclass[twocolumn]{aastex7}

\usepackage{savesym}
\usepackage{graphicx}	
\usepackage{amsmath}	
\usepackage{amssymb}	

\renewcommand{\figurename}{Fig.}

\savesymbol{longtable}
\usepackage{xcolor}

\usepackage[utf8]{inputenc}
\usepackage[english]{babel}

\restoresymbol{SUB}{longtable}

\usepackage{physics}

\begin{document}

\title{MeV cosmic-ray electrons modify the TeV pair-beam plasma instability}

\author[0009-0008-0835-2795]{Mahmoud Alawashra}
\affiliation{Deutsches Elektronen-Synchrotron DESY, Platanenallee 6, 15738 Zeuthen, Germany}
\email{mahmoud.al-awashra@desy.de}

\author[0000-0001-8849-4866]{Yuanyuan Yang}
\affiliation{Center for Cosmology and Astroparticle Physics, 191 West Woodruff Avenue, Columbus, OH 43210, USA}
\affiliation{Department of Astronomy, The Ohio State University, 140 West 18th Avenue, Columbus, OH 43210, USA}
\affiliation{Department of Physics, Northeastern University, 440 Huntington Avenue, Boston, MA 02115, USA}
\email{yuanyuan.yang@cfa.harvard.edu}

\author[0000-0002-2951-4932]{Christopher M. Hirata}
\affiliation{Center for Cosmology and Astroparticle Physics, 191 West Woodruff Avenue, Columbus, OH 43210, USA}
\affiliation{Department of Astronomy, The Ohio State University, 140 West 18th Avenue, Columbus, OH 43210, USA}
\affiliation{Department of Physics, The Ohio State University, 191 West Woodruff Avenue, Columbus, OH 43210, USA}
\email{hirata.10@osu.edu}

\author[0000-0001-8290-5417]{Heyang Long}
\affiliation{Center for Cosmology and Astroparticle Physics, 191 West Woodruff Avenue, Columbus, OH 43210, USA}
\affiliation{Department of Physics, The Ohio State University, 191 West Woodruff Avenue, Columbus, OH 43210, USA}
\email{long.1697@buckeyemail.osu.edu}

\author[0000-0001-7861-1707]{Martin Pohl}
\affiliation{Deutsches Elektronen-Synchrotron DESY, Platanenallee 6, 15738 Zeuthen, Germany}
\affiliation{Institute for Physics and Astronomy, University of Potsdam, D-14476 Potsdam, Germany}
\email{martin.pohl@desy.de}

\received{-}
\revised{-}
\accepted{-}
\published{-}
\submitjournal{The Astrophysical Journal}

\begin{abstract}
Relativistic pair beams created in the intergalactic medium (IGM) by TeV gamma rays from blazars are expected to produce a detectable GeV-scale electromagnetic cascade, but the cascade component is absent in the spectra of many hard-spectrum TeV-emitting blazars. One common explanation is that weak intergalactic magnetic fields deflect the electron-positron pairs away from our line of sight. An alternative possibility is that electrostatic beam-plasma instabilities drain the energy of these pairs before a cascade
can develop. Recent studies have shown that beam scattering by oblique electrostatic modes leads to minimal energy loss. But these modes might be suppressed by linear Landau damping (LLD) due to MeV-scale cosmic-ray electrons in the IGM. In this work, we explore the impact of LLD on the energy-loss efficiency of plasma instabilities in pair beams associated with 1ES 0229+200. We find that LLD effectively suppresses oblique electrostatic modes, while quasi-parallel ones grow to larger amplitudes. In this way LLD enhances the energy-loss efficiency of the instability by more than an order of magnitude.

\end{abstract}




\section{Introduction}

The vast majority of the universe's volume is occupied by the intergalactic medium (IGM). The effects of various components within the IGM (such as radiation fields, thermal plasma, and magnetic fields) on the propagation of high-energy gamma rays from extragalactic sources like blazars have been widely explored in the literature \citep{2010Sci...328...73N,Broderick_2012}. In addition to these constituents of the IGM, recent study have revealed the existence of a population of MeV cosmic-ray electrons, generated through Compton scattering of MeV gamma rays off the IGM’s thermal electrons \citep{Yang_2024}. In this paper, we investigate, for the first time, the impact of these IGM cosmic-ray electrons on the propagation of high-energy gamma rays from blazars.

Blazars are active galactic nuclei (AGN) with their relativistic jet pointing toward Earth. Blazars are considered to be among the most powerful extragalactic sources of high-energy radiation, emitting TeV gamma rays that interact with the extragalactic background light (EBL) to produce relativistic electron–positron pair beams in the intergalactic medium (IGM) via photon-photon pair production. These pair beams are expected to generate a GeV-scale inverse-Compton cascade as they upscatter cosmic microwave background (CMB) photons. However, observations of hard-spectrum TeV-emitting blazars, such as 1ES 0229+200, show a surprising absence of this secondary GeV component, suggesting that additional physical mechanisms affect the propagation of the pair beams.

One commonly proposed explanation for the missing GeV cascade is the presence of weak intergalactic magnetic fields (IGMFs), which could deflect pair beams enough to prevent their emission from reaching Earth or introduce significant time delays \citep{2010Sci...328...73N}. Observations of distant blazars from Fermi-LAT and ground-based telescopes have been used to find a lower limit on the strength of the IGMFs that is around $7 \times 10^{-16}$ Gauss for a homogeneous IGMF and a blazar duty cycle of ten years \citep{Aharonian_2023}. For longer blazar duty cycles, stronger IGMFs are needed. If the IGMF is strong enough ($\gtrsim 10^{-14}$ G), the $e^\pm$ will be strongly deflected and the superposition of GeV cascades from all of the blazar-like AGN (including those not pointed at Earth) will appear in the isotropic gamma ray background. Modeling the contributions of star-forming galaxies and AGNi can account for essentially all of the observed background \citep{2023JCAP...02..003B}, leaving little room for an additional contribution from isotropized cascades \citep{2023arXiv230301524B}. 

An alternative scenario is that electrostatic beam-plasma instabilities may efficiently extract energy from the beam before it can develop a cascade. In this case, interactions between the relativistic pairs and the background IGM plasma could dissipate a significant fraction of the beam energy as a  heating of  the IGM rather than re-emitting it as scandary gamma-ray photons \citep{Broderick_2012,Schlickeiser_2012,Schlickeiser_2013,2014ApJ...790..137B,Sironi_2014,Chang_2014,Supsar_2014,2016ApJ...833..118C,2016A&A...585A.132K,2017A&A...607A.112R,Vafin_2018,AlvesBatista:2019ipr,shalaby_broderick_chang_pfrommer_puchwein_lamberts_2020}. 

It was argued that the efficacy of such instabilities is limited either by the IGM plasma inhomogeneity that induce loss of resonance between the beam particles and the plasma oscillations or by nonlinear Landau scattering on the IGM ions that transports wave energy out of resonance and to smaller wave numbers \citep{Miniati_2013,Sironi_2014,Vafin_2019}.

Recent numerical studies have shown that the dominant contributor to pair-beam energy dissipation, oblique electrostatic modes, primarily broaden the initially tiny beam spread, $\Delta \theta \sim 1/\gamma$, by one to two order of magnitudes, depending on the distance from the blazar and the Lorentz factor of the pairs, which suppresses the instability growth yielding only minimal beam energy loss \citep{Perry_2021,Alawashra_2024,Alawashra_2025}. 

The oblique modes may themselves be subject to Linear Landau Damping (LLD) by the MeV-scale cosmic-ray electrons in the IGM \citep{Yang_2024}. Those cosmic ray electrons are produced from the MeV gamma-ray background photons Compton scattering off the thermal electrons in the IGM; since the Universe is transparent to MeV-scale gamma rays, this background is quasi-isotropic (coming from many sources spread across the Hubble volume), and so are the resulting cosmic-ray electrons. The LLD rates are much faster than the linear instability growth rates and the collisional damping rate at the oblique instability modes \citep{Yang_2024}. LLD requires resonant particles ($v = v_{\rm ph} \cos\vartheta\simeq \omega/k \cos\vartheta$, where $v$ is the velocity of the particle, $v_{\rm ph}$ is the phase velocity of the wave, and $\vartheta$ is the angle between the particle velocity and wave vector). Therefore, electrostatic modes with very small angles relative to the pair beam, and hence minimal wave number, $k$, can survive: these modes resonate with the ultrarelativistic pair beam, but they ``outrun'' the cosmic rays. 

In this work, we investigate the impact of LLD on the nonlinear feedback of the plasma instability on the TeV-induced pair beams, with a particular focus on 1ES 0229+200. By incorporating LLD into the instability feedback calculation, we investigate the LLD suppression of the oblique electrostatic modes and the growth of the quasi-parallel ones. The quasi-parallel modes allow for higher energy transfer from the beam to the plasma oscillations. We systematically investigate the impact of this new beam-plasma dynamic with LLD on the energy-loss efficiency of plasma instability. We compare the beam energy loss rates due to the instability for the cases of including/omitting the LLD.

The structure of this paper is as follows. We start by introducing the quasi-linear evolution description of the beam-plasma resonant modes in section \ref{sec:W} including the Linear Landau damping by IGM cosmic-ray electrons in subsection \ref{sec:LLD}. In section \ref{sec:f}, we present the pair beam transport equation in the IGM. The numerical setup for the numerical integration of the temporal equations of the beam-plasma system is presented in section \ref{sec:4}. The results are presented and discussed in section \ref{sec:5}. Finally, we conclude in Section \ref{sec:con}.

\section{Beam-plasma resonant modes temporal evolution}\label{sec:W}

The predominant instability of the blazar-induced pair beam is the electrostatic resonant instability. The beam particles with velocity, $\boldsymbol{v}$, excite Langmuir oscillations with wave vector, $\boldsymbol{k}$, under the resonance condition \citep{Brejzman_1974}
\begin{equation}
    \omega_p -\boldsymbol{k}\cdot\boldsymbol{v} =0,
\end{equation}
where $\omega_p = (4\pi n_e e^2/m_e)^{1/2}$ is the plasma frequency of the IGM background plasma with density $n_e$. 

The blazar-induced pair beams are produced with Lorentz factors, $\gamma$, between $10^4$ and $10^7$ and their angular spread is very narrow, $\Delta \theta \sim \gamma^{-1}$. The corresponding resonant oscillations are excited with angles of wide range with the beam axis, $\theta' \sim 10^{-6} - 1$ radian. Their parallel wave number component, $k_{||}$, is very narrow and stretches over the range,  $\left({ck_{||}}/{\omega_p}-1\right) \sim 10^{-13} - 10^{-7}$, with the smallest one corresponding to the parallel angles and the largest once to the oblique angles.

The amplitude of the resonant oscillations grows exponentially in the linear phase of the instability. This linear phase is followed by a nonlinear phase in which the electric-field fluctuations feed back on the beam, scattering it in the transfer direction. The broadening of the beam reduces the initially large linear growth rate. Then the Coulomb collisions of the oscillating IGM electrons with the quasi-static ions become significant as well. 

Beside the thermal plasma in the IGM, there is a cosmic-ray population of quasi-relativistic electrons due to the Compton scattering of MeV gamma-ray photons with thermal IGM electrons \citep{Yang_2024}. Those isotropic cosmic-ray electrons cause linear Landau damping of the Langmuir waves in the IGM. In fact, the damping rates, $ \omega_\mathrm{LLD,i}(k)$, are much faster than the linear growth rate of the instability at oblique angles, $\theta' \sim 1$. We describe the LLD in more detail in section \ref{sec:LLD}.
 
Including the linear Landau damping caused by the Compton-induced electrons, and using the quasilinear approximation, we may build a kinetic equation to follow the spectrum of electric field fluctuations\footnote{Defined as the electric energy per unit volume in real space, per unit volume in $\boldsymbol k$-space; note that the total oscillation energy is twice the electric energy since there is also kinetic energy of the oscillating electrons}, $W$, as a function of time, $t$. The temporal evolution is well described by the following equation 
\begin{equation}\label{eq:W}
    \frac{\partial W (\vb{k},t)}{\partial t} = 2 (\omega_i (\vb{k},t) +  \omega_\mathrm{LLD,i}(k) + \omega_c(k)) W(\vb{k}).
\end{equation}
The quasilinear description is valid as long as the energy density of the electric-field fluctuations is much less than the thermal energy density of the background plasma. 

The first term on the right hand side of equation \ref{eq:W} is the linear growth rate of the electrostatic instability \citep{Brejzman_1974},
\begin{equation}\label{eq:wi}
    \omega_i(\boldsymbol{k},t) = \omega_p \frac{2\pi^2 e^2}{k^2} \int d^3\boldsymbol{p} \left(\boldsymbol{k}\cdot\frac{\partial f(\boldsymbol{p},t)}{\partial \boldsymbol{p}}\right) \delta(\omega_p - \boldsymbol{k}\cdot\boldsymbol{v}),
\end{equation}
where $f$ is the beam momentum distribution whose temporal evolution we cover in section \ref{sec:f}. The third one is the Coulomb collisional damping rate \citep{Tigik_2019}
\begin{equation}\label{eq:colldamp}
     \omega_c(k) = - \omega_p \frac{g}{6\pi^{3/2}}\frac{1}{(1+3k^2\lambda_D^2)^3},
\end{equation}
where $g=(n_e\lambda_D^3)^{-1}$ is the plasma parameter, $\lambda_D = 6.9 \text{ cm} \sqrt{\frac{T_e/ K}{n_e/\text{cm}^{-3}}} $ is the Debye length, $n_{e} = 10^{-7} (1+z)^3\, \text{cm}^{-3}$ is the density of IGM electrons, fiducially taken as half of the mean density. We shall discuss the impact of the choice of density later in the paper. Finally, $T_e = 10^{4}\, \text K$ is their temperature.

The total electric field energy density in the electrostatic fluctuations is given as
\begin{equation}\label{eq:Wtot}
    W_{\text{tot}} = 2 \pi \int dk_\perp k_\perp \int dk_{||} W(k_\perp,k_{||}),
\end{equation}
and the differential energy loss of the beam due to the growth of the electrostatic mode is given by \citep{Vafin_2018}
\begin{equation} \label{eq:dUdt}
\begin{split}
    \frac{dU_b}{dt} = & - 2 \left[\frac{dW_{\text{tot}}}{dt} \right]_{\omega_i} \\ = & -8\pi\int dk_\perp k_\perp \int dk_{||} W(k_\perp,k_{||}) \omega_{i}(k_\perp,k_{||}),
\end{split}
\end{equation}
where the total beam energy density is defined as
\begin{equation}
    U_b = 2 \pi \int dp p^2 \int d\theta \sin{\theta} m_e c^2 \gamma f(p,\theta).
\end{equation}

In the rest of this section, we explain in detail the second term on the right hand side of equation \ref{eq:W} representing the linear Landau damping caused by the cosmic ray electrons in the intergalactic medium.

\subsection{Linear Landau damping by IGM cosmic-rays electrons}\label{sec:LLD}

Plasma waves in the intergalactic medium (IGM) are subject to various damping mechanisms. One of the most fundamental is Landau damping, a collisionless process first described by \cite{landau}. In a cold plasma (where the electron thermal velocity is negligible compared to the wave's phase velocity), no Landau damping occurs because there are no electrons resonant with the wave. However, in a warm or hot plasma (finite electron temperature), the electron velocity distribution is broad enough that some electrons move at velocities close to the wave's phase velocity. These electrons can then resonate with the wave, and the key outcome is that the wave frequency acquires a real part, $ \omega_\mathrm{LLD,r}(k)$, corresponding to the oscillation frequency, and an imaginary part, $ \omega_\mathrm{LLD,i}(k)$, corresponding to the damping (if $ \omega_\mathrm{LLD,i}(k) < 0$) or growth (if $ \omega_\mathrm{LLD,i}(k) > 0$) of the wave. For small wavenumbers, $k$, with phase velocities $\omega_p/k\gg \sqrt{k_{\rm B}T_e/m_e}$, the Landau damping from thermal electrons is strongly suppressed, but Landau damping from non-thermal electrons (in our case, the MeV electrons) may still proceed. 

When the amplitude of the electrostatic wave (e.g., a Langmuir wave) is sufficiently small, we can work in the linear regime: the wave does not significantly modify the velocity distribution of the background electrons. In this work, we follow the approach of \cite{1995PhRvL..74.2232B} and \cite{Yang_2024} to compute such linear Landau damping rate (LDD) of Langmuir waves in the IGM. We account for an isotropic population of ~MeV cosmic-ray electrons distribution as modeled in \cite{Yang_2024}.

In particular, \cite{Yang_2024} derive an expression (their Equation 34) for the imaginary correction to the plasma frequency, which we denote $ \omega_\mathrm{LLD,i}(k)$. That expression reads schematically:
\begin{equation}
\begin{split}
 \omega_\mathrm{LLD,i} (\vb{k}) = &\frac{\pi\omega_{\rm p}}{2n_e} \left( \frac{m_e\omega_{\rm p}}{k} \right)^2 \int_{{\mathbb R}^3} d^3{\bf p}\,
\,\delta \left(m_e\omega_{\rm p} - \frac{{\bf k}\cdot{\bf p}}\gamma\right) \\ & \times \left(
\,{\bf k}\cdot\nabla_{\bf p}f_e({\bf p}) \right),
\label{eq:Im-omega}
\end{split}
\end{equation}
where $f_e({\bf p})$ is the electron momentum distribution and ${\bf p}$ is the electron momentum. While LLD is often derived assuming an analytic form for $f_e({\bf p})$ (e.g., a Maxwellian), \cite{Yang_2024} instead numerically model the cosmic-ray electron distribution. For an isotropic distribution, one can perform the angular integration in equation \ref{eq:Im-omega} analytically to find
(\cite{Yang_2024}, Eq. 37)
\begin{equation}
\begin{split}
 \omega_\mathrm{LLD,i}(k)
= &-\frac\pi 4 \omega_{\rm p} \left( \frac{\omega_{\rm p}}{ck} \right)^3 \frac{m_ec^2}{n_{e}} \\ & \times \Bigg[ \frac{E_{\rm min}+m_ec^2}{\sqrt{E_{\rm min}(E_{\rm min}+2m_ec^2)}} N(E_{\rm min}) \\ & + 2\int_{E_{\rm min}}^\infty \frac{N(E)\,dE}{\sqrt{E(E+2m_ec^2)}}
 \Theta\left( \frac{ck}{\omega_{\rm p}}- 1\right) \Bigg],
\label{eq:ImCR}
\end{split}
\end{equation}
where $N(E)$ is the number of electrons per unit physical volume per unit energy [units: $\rm cm^{-3}\ erg^{-1}$] modeled by \cite{Yang_2024}, $\Theta$ is the Heaviside step function, and $E_{\rm min}$ is the kinetic-energy threshold corresponding to electrons whose velocity matches the wave's phase velocity, $v_\phi = \omega_p/v$ \citep{Yang_2024}
\begin{equation}\label{eq:Emin}
    E_{\rm min} = m_e c^2 [ (1-\omega_p^2/c^2k^2)^{-1/2}-1].
\end{equation}
Physically, electrons with energies above $E_{\rm min}$ can resonate with the oscillation with  wave number $k$, and contribute to the damping.

\begin{figure}
    \centering
    \includegraphics[width=0.5\textwidth]{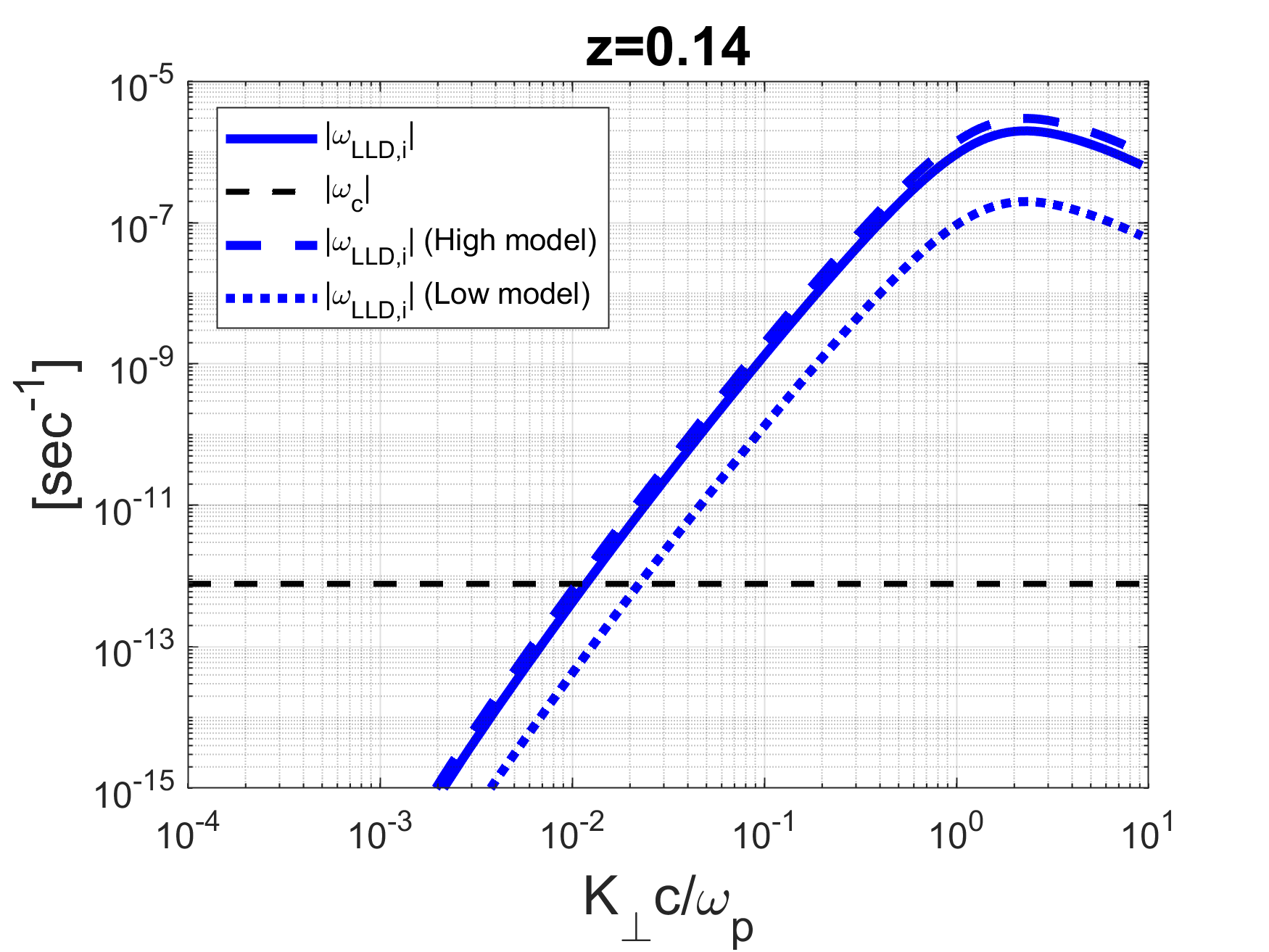}
    \caption{ The modulus of the LLD rate due to MeV cosmic-ray electrons in the IGM, $|\omega_{\text{LLD},i}|$, as a function of the normalized perpendicular wave number, $c k_\perp / \omega_p$, with a strictly parallel wave number, $c k_{||}/\omega_p = 1$, at redshift $z = 0.14$, relevant to the blazar 1ES 0229+200. The damping rate is computed for three different MeV gamma-ray background models: the fiducial model used in \cite{Yang_2024} (solid blue), a high-background scenario with a scaling factor of 1.5 (dashed blue), and a low-background scenario with a scaling factor of 0.1 (dotted blue). The black dashed line represents the absolute value of the collisional damping rate, $|\omega_c|$, given by equation \ref{eq:colldamp}. 
        \label{fig:LDrate}
    }
\end{figure}

The calculations of the cosmic-ray electron distribution in \cite{Yang_2024} incorporate a source term based on the fiducial gamma-ray background model (“Q18”) in \citet{10.1093/mnras/stz174}. To account for the uncertainties in this background model, we introduce scaling factors that adjust the gamma-ray flux within reasonable observational constraints. For an upper-limit scenario, we adopt a "High" model where the gamma-ray background is increased by a factor of 1.5. This choice is justified as it places the model near the upper error bounds of the COMPTEL and EGRET measurements in the 1–100 MeV range (see Figure 6 in \citet{10.1093/mnras/stz174}).

Conversely, for a lower-limit scenario, we define a "Low" model by considering only the contribution from star-forming galaxies, as predicted in \cite{Lacki_2014}. This is a conservative approach that excludes contributions from active galactic nuclei (AGN), whose spectra remain poorly constrained in the relevant energy range. The model in \citet{Lacki_2014} predicts a gamma-ray flux that is lower by a factor of 0.1–0.8 at the energy $E_\gamma$ = 300 MeV, and 0.03–0.5 around $E_\gamma$ = 30 MeV. Given that electrons must satisfy $\gamma > 1/(c k_\perp/\omega_p)$ (cf. equation \ref{eq:Emin}) to contribute to the damping, the higher-energy portion of the spectrum is most relevant to our analysis for perpendicular wave numbers, $ck_\perp/\omega_p \sim 10^{-4} - 10^{-2}$. Thus, we adopt a scaling factor of 0.1 as a representative "Low" model.

By implementing these two extreme cases, we effectively bracket the possible variations in the gamma-ray background and their potential impact on our results. This approach provides a robust framework for assessing the sensitivity of our model to uncertainties in the extragalactic gamma-ray flux.

We present in figure \ref{fig:LDrate} the modulus of the linear Landau damping rate caused by the MeV cosmic-ray electrons, $ \omega_\mathrm{LLD,r}$, as a function of the normalized perpendicular wave number, $c k_\perp/\omega_p$, with strictly parallel wave number, $ck_{||}/\omega_p = 1$, and at red shift $z = 0.14$ relevant to the blazar 1ES 0229+200. The damping rate is computed for three different MeV gamma-ray background models: the fiducial model used in \cite{Yang_2024} (solid blue), a high-background scenario with a scaling factor of 1.5 (dashed blue), and a low-background scenario with a scaling factor of 0.1 (dotted blue). For reference, we also include the absolute value of the collsional damping rate, $|\omega_c|$, given by equation \ref{eq:colldamp} (black dashed).

We see in figure \ref{fig:LDrate} that the peak of the LLD rates occurs at the oblique angles with the beam axis, near $c k_\perp/\omega_p \sim 2$. Notably, at normalized perpendicular wave numbers below $10^{-2}$, the LLD rate falls below the collisional damping rate implying that oscillations in this regime can still grow as the Linear growth rates are much higher than the collisional damping rate. At $c k_\perp/\omega_p \approx 1$, the LLD rate exceeds the initial linear growth rate of the electrostatic pair-beam instability. This suggest that LLD could play a significant role in the evolution of the instability as those oscillation mode are responsible for the scattering of the beam particles \citep{Perry_2021,Alawashra_2024,Alawashra_2025}.

\section{Pair Beam evolution}\label{sec:f}

In this section, we present the theoretical framework governing the temporal evolution of the pair beam within the IGM. The key processes driving this evolution include beam diffusion induced by plasma instability feedback, pair production, and inverse-Compton cooling. These processes are encapsulated in the following transport equation
\begin{equation}\label{eq:f}
\begin{split}
     \frac{\partial f(p,\theta)}{\partial t} = &  \frac{1}{p^2\theta}\frac{\partial}{\partial \theta}\left(\theta D_{\theta\theta}(\theta,t) \frac{\partial f(p,\theta)}{\partial \theta}\right) \\ & + \frac{1}{p^2}\frac{\partial}{\partial p}\left(- \Dot{p}_{\text{IC}} p^2 f(p,\theta)\right) + Q_{ee}(p,\theta).
\end{split}
\end{equation}

The first term on the right-hand side accounts for the feedback of the excited electrostatic oscillations on the beam, modeled using a Fokker-Planck diffusion approach \citep{Brejzman_1974}. We neglect momentum diffusion from instability feedback, as the angular $\theta \theta$ diffusion dominates the beam’s response to the instability \citep{Perry_2021,Alawashra_2024}.

The diffusion coefficient, $D_{\theta \theta}$, is given by an integral over the oscillating electric field energy density, $W$, involving the wave-particle resonance condition \citep{1971JETP...32.1134R}
\begin{equation}\label{eq:D}
    D_{\theta \theta}(\boldsymbol{p},t) = \pi e^2 \int d^3\boldsymbol{k} W(\boldsymbol{k},t) \frac{k_\theta k_\theta}{k^2} \delta(\boldsymbol{k}\cdot\boldsymbol{v}-\omega_p),
\end{equation}
where $e$ is the electric charge. The wave vector, $\boldsymbol{k}$, is expressed in spherical coordinates, $(k,\theta',\varphi')$, with the beam axis aligned along the z-direction. The projection of the wave vector onto the angular spatial component of the beam ($\theta$-direction) is given by
\begin{equation}
    k_\theta = \boldsymbol{k}\cdot\boldsymbol{\hat{\theta}} = k[\sin{\theta'}\cos{\theta}\cos{\varphi'}-\cos{\theta'}\sin{\theta}].
\end{equation}

The second term represents energy loss due to inverse Compton cooling with the cosmic microwave background (CMB) photons. Since the Lorentz factors of the produced pairs fall within the Thomson scattering regime, where $(4\epsilon_{\text{CMB}} \gamma \ll m_e c^2)$, the cooling rate is given by
\begin{equation}
    \Dot{p}_{\text{IC}} = - \frac{4}{3} \sigma_\text{T} u_\text{CMB} \gamma^2,
\end{equation}
where $\sigma_\text{T}$ is the Thompson cross-section, and $u_\text{CMB} = u_\text{CMB,0}(1+z)^4$ is the CMB energy density that scales with redshift, $z$. We use the value of the CMB energy density of at redshift zero to be 0.26 eV cm$^{-3}$ \citep{Fixsen_2009}. The third term, $Q_{ee}(p,\theta)$, represents the source term for pair production due to gamma-ray annihilation with the EBL photons. We employ the source terms derived in \citet{Alawashra_2025} for the pair beams produced at different distances in the IGM from the blazar 1ES 0229+200. 


The system of equations \ref{eq:f} and \ref{eq:W} self-consistently describes the temporal evolution of the beam-plasma system, incorporating both beam production and cooling. Plasma oscillations influence the beam through the diffusion coefficient, $D_{\theta\theta}$ (equation \ref{eq:D}), while the beam, in turn, modulates the oscillation growth via the linear growth rate, $\omega_i$ (equation \ref{eq:wi}). We do not incorporate here the beam energy loss to the instability growth, but we can calculate the differential total energy transfer from the beam to the plasma oscillations at any given time using equation \ref{eq:dUdt}.

\section{Numerical setup}\label{sec:4}

The coupled equations \ref{eq:f} and \ref{eq:W} are numerically integrated over time. To analyze the impact of the LLD term on beam energy loss due to instability, we perform the calculation with and without the LLD term in equation \ref{eq:W}. The numerical setup remains otherwise identical in both scenarios.

The numerical integration of equations \ref{eq:f} and \ref{eq:W} demands time steps considerably shorter than the inverse-Compton cooling times of the pairs, owing to the substantial growth and damping rates involved. However, pair beams propagate over cosmological distances, covering timescales of billions of years. To handle this significant scale separation, we consider beam production rates, $Q_{ee}$, at logarithmically binned distances 
from a 1ES 0229+200-like source \citep{Alawashra_2025}, because that is slowly evolving. These production rates serve as input for the beam-plasma calculations, enabling us to determine the energy loss rate of the beams using equation \ref{eq:dUdt}, once equilibrium is reached.

We have evaluated the LLD rate at redshift of the 1ES 0229+200 source, $z=0.14$, using the publicly available code from \cite{Yang_2024}. This code internally handles the detailed numerical model for the IGM cosmic ray electrons distribution, $N(E)$ (or equivalently $f_e({\bf p})$), and performs the required integrals to yield the LLD rate (equation \ref{eq:ImCR}). We calculate the LLD rate on the same 2D wave-number grid as is used to numerically integrate equation \ref{eq:W}. Three different LLD levels are considered, based on three different MeV gamma-ray background models as we demonstrated in section \ref{sec:LLD}: the fiducial model used in \cite{Yang_2024}, a high-background scenario with a scaling factor of 1.5, and a low-background scenario with a scaling factor of 0.1.

We numerically solve the beam transport equation (\ref{eq:f}) with the initial condition of the beam accumulated over a year, shorter than the initial instability growth time, $\omega^{-1}_{i,\text{max}}\sim 1 - 100$ yr, and the expected inverse-Compton cooling time, $\tau_{\text{IC}}\sim 2.5$ Myr$ \times {10^6}/{\gamma}$. We adopt operator splitting of equation \ref{eq:f}, solving the beam cooling and injection terms using the Crank-Nicolson scheme with forward momentum derivative \citep{10.1046/j.1365-8711.1999.02538.x}, while the diffusive feedback of the instability is treated using the Crank-Nicolson scheme with central angular derivative \citep{Alawashra2024}. 

\begin{figure}
\centering
        \includegraphics[width=\columnwidth]{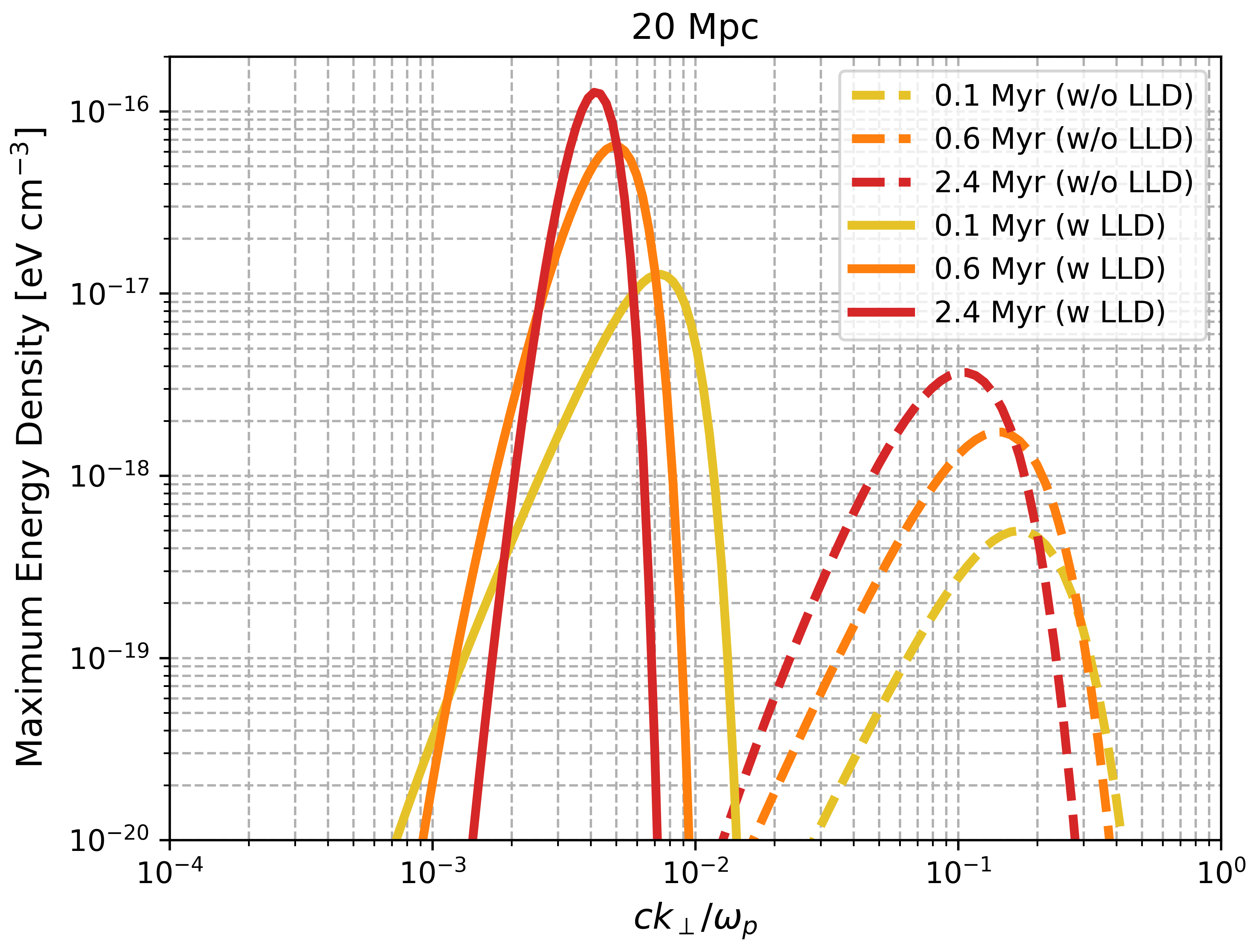}
        \caption{The maximum energy density, $2 \pi k_\perp \Delta k_\perp \left(\Delta k_{||} W(k_\perp, k_{||}) \right)_{\text{max}},$ of electrostatic modes at distance 20 Mpc from the source, plotted as a function of the perpendicular wave number, $ck_\perp/\omega_p$, with color indicating time. Dashed curves follow the time evolution in the absence of LLD, whereas solid lines include the effect of LLD. }
        \label{fig:Wmax}
\end{figure}

The wave balance equation, \ref{eq:W}, is solved using the forward time scheme after transforming to the time derivative of the natural logarithm of the electric-field energy density, $W$. The initial oscillation energy density corresponds to the thermal energy density of the IGM background plasma \citep{Vafin_2019}. We employ an adaptive time-stepping method, selecting the smallest value among $ \omega_\mathrm{LLD,\text{max}}^{-1}$, $\omega_{i,\text{max}}^{-1}$, $\omega_c^{-1}$, and the inverse of the most rapid rate of change in the beam distribution. A universal maximum time step of $10^{11}$ seconds is enforced. To ensure numerical stability, we validate our approach using time steps reduced by two orders of magnitude. 

The coordinate system follows that used by \citet{Alawashra_2024} to accurately resolve the narrow oscillation spectrum, $(ck_\perp/\omega_p,\theta^R)$, where $\theta^R = (ck_{||}/\omega_p-1)/(ck_\perp/\omega_p)$. Here, $k_\perp$ and $k_{||}$ denote the perpendicular and parallel components of the wave vector relative to the beam propagation direction, respectively. We used a logarithmic binning of the normalized perpendicular wave number, $ck_\perp/\omega_p$, with the range between $10^{-3}$ and $10^{1}$ when omitting the LLD and with the range between $10^{-4}$ and $2\times 10^{-2}$ when including the LLD.

\begin{figure}
    \centering
    \includegraphics[width=0.5\textwidth]{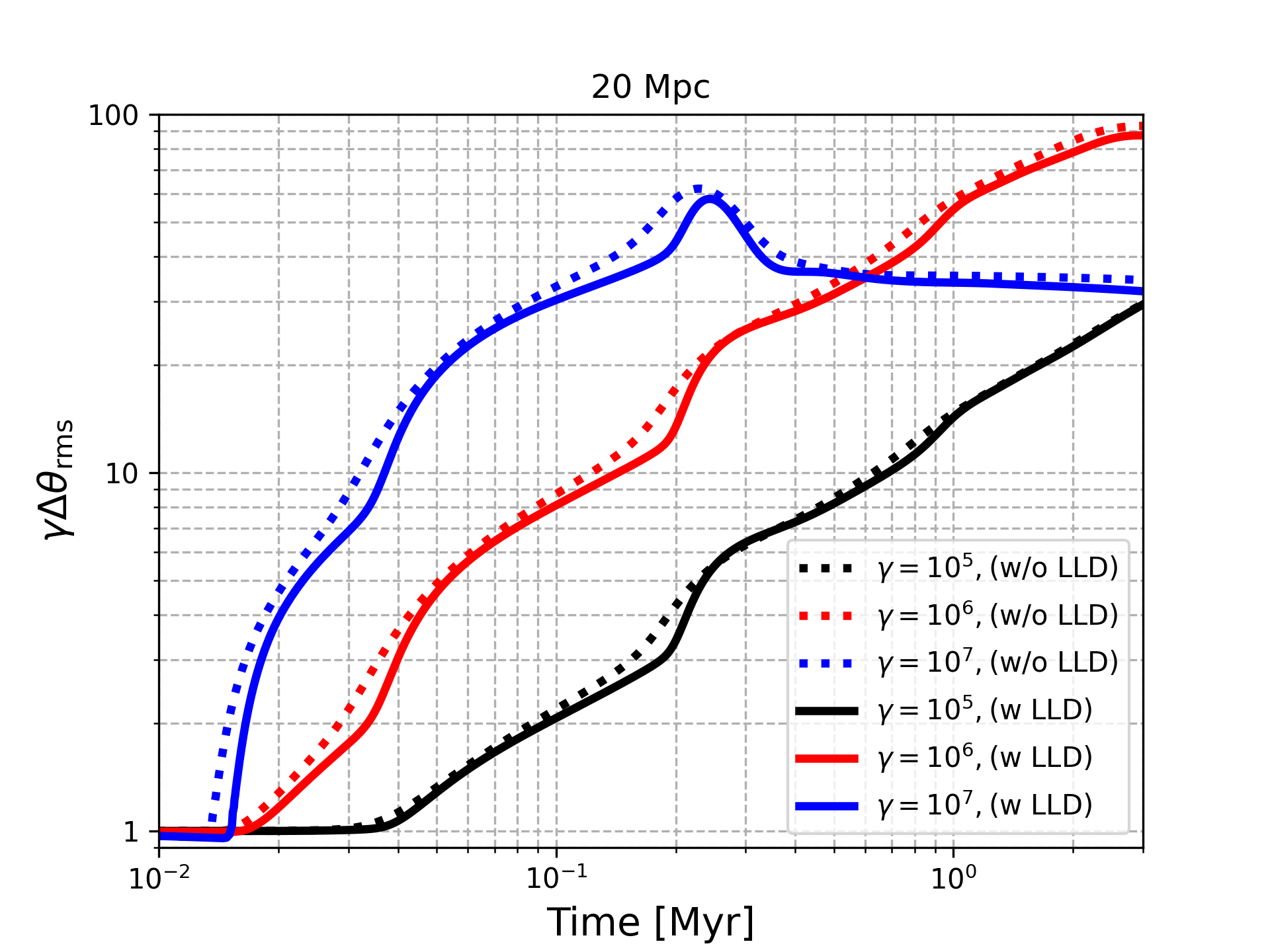}
    \caption{The time evolution of the beam angular spread (root mean square) for different Lorentz factors at the distance of 20 Mpc from the source. Dotted curves are for the simulation in the absence of LLD, where solid lines follow the one including the LLD.
        \label{fig:delta_theta}
    }
\end{figure}

\section{Results and discussion} \label{sec:5}

In order to asses the impact of the LLD on the evolution of the instability of the blazar-induced pair beams, we have performed the numerical calculations of the quasilinear feedback (equations \ref{eq:f} and \ref{eq:W}) in two setups, with and without the LLD term in equation \ref{eq:W}. The LLD rate due to MeV-scale cosmic-ray electrons in the IGM is illustrated in figure \ref{fig:LDrate} as a function of the perpendicular wave number. We expect from figure \ref{fig:LDrate} that the moderately oblique electrostatic  modes ($ck_{\perp}/\omega_p \sim 10^{-2}$–$1$), will be effectively suppressed by LLD. In contrast, quasi-parallel modes with $ck_{\perp}/\omega_p \lesssim 10^{-2}$ will remain largely unaffected. 

\begin{figure}
    \centering
    \begin{minipage}[b]{0.47\textwidth}
        \centering
        \includegraphics[width=\textwidth]{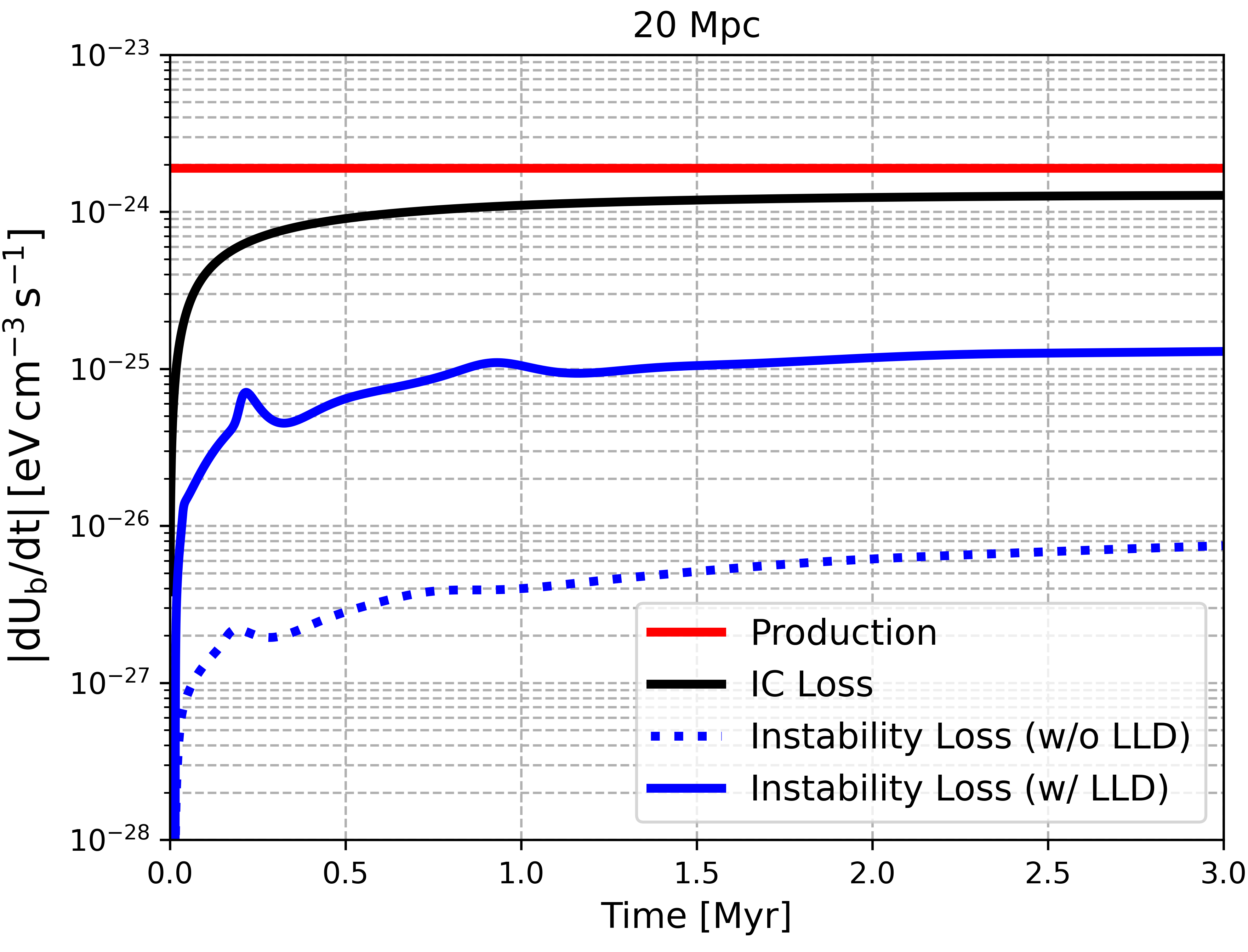} 
    \end{minipage}
    \hfill
    \begin{minipage}[b]{0.47\textwidth}
        \centering
        \includegraphics[width=\textwidth]{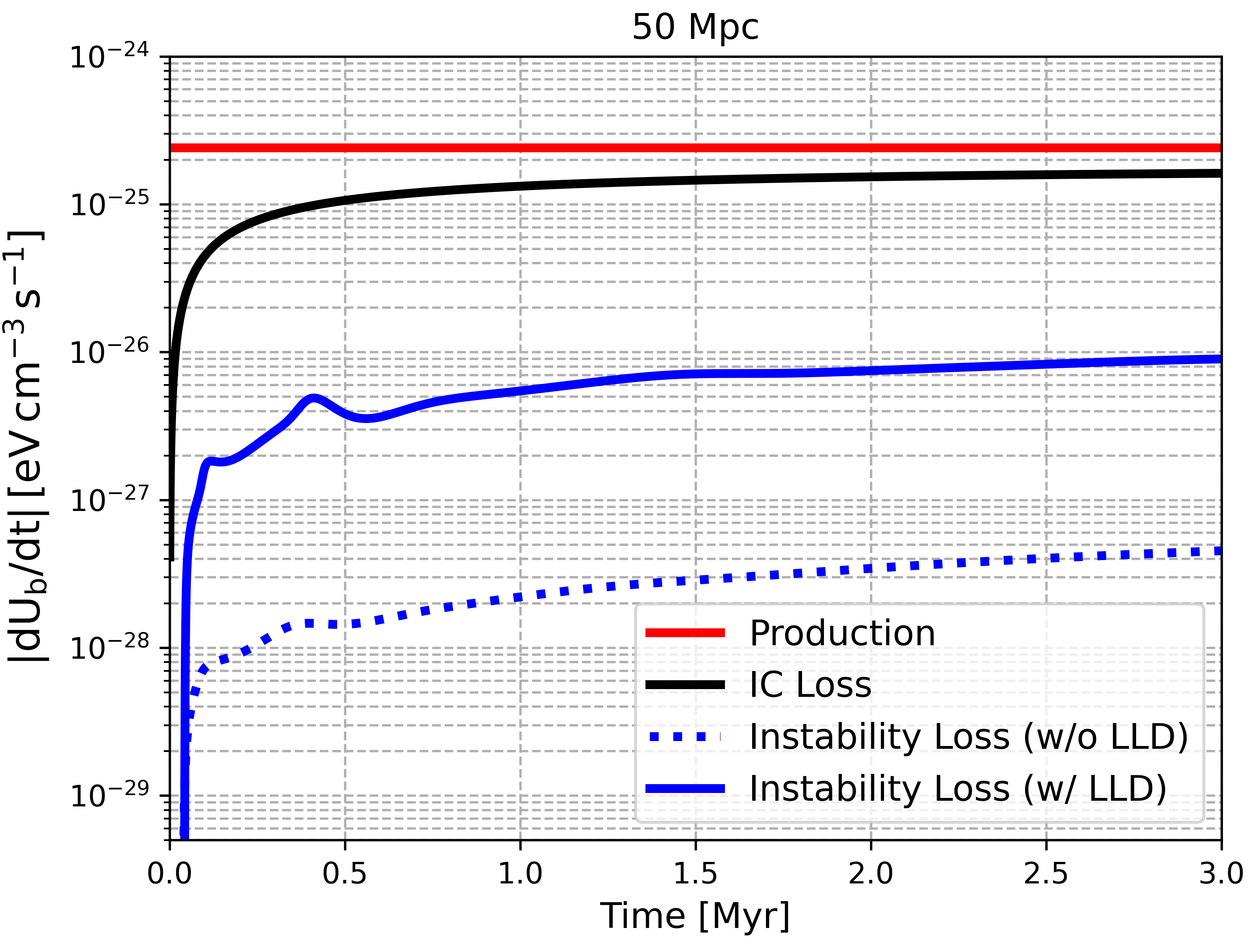} 
    \end{minipage}
    \caption{Time evolution of the absolute rate of change of the beam energy, $|dU_b/dt|$, at distances of 20 and 50 Mpc from the source. The red line represents the pair production rate, while the black line indicates energy loss due to inverse Compton (IC) cooling. The blue dotted curve shows the instability-induced energy loss without LLD, whereas the solid blue curve represents the instability energy loss with LLD included. The presence of LLD significantly enhances the instability energy loss, leading to a more efficient dissipation of beam energy into plasma oscillations.}
    \label{fig:dUdt}
\end{figure}

We perform the quasilinear calculations at various distances from a 1ES 0229+200-like source ranging from 5 to 125 Mpc. We stop the calculation after $3~$Myr when a steady state of the beam and the oscillation is achieved. The results presented in the figures \ref{fig:Wmax}, \ref{fig:delta_theta} and \ref{fig:dUdt} are derived for the LLD rates based on the fiducial MeV gamma-ray background model used in \cite{Yang_2024} (solid blue in \ref{fig:LDrate}).           

Figure \ref{fig:Wmax} demonstrates the impact of the LLD on the evolution of unstable modes. It shows the maximum electric-field energy density as a function of the perpendicular wave number, $ck_{\perp}/\omega_p$, at a representative distance of 20 Mpc from the source. In the absence of LLD (dashed curves), a high intensity is observed for modes around $ck_{\perp}/\omega_p \approx 0.1$ that are effective at scattering beam particles transversely. When LLD is included (solid curves), those oblique oscillations are strongly suppressed, and only the low-$k_{\perp}$, quasi-parallel modes survive. As a result of LLD, the wave intensity is preferentially carried by small-angle modes. 

Notably, figure \ref{fig:Wmax} also shows that with LLD the surviving modes carry a higher energy density than they would without LLD. This arises because the instability feedback on the beam (via quasilinear diffusion) is much less efficient for small $k_{\perp}$ modes: the angular diffusion coefficient scales as $D_{\theta\theta}\propto k_{\perp}^4$ (see equation \ref{eq:D}), whereas the wave energy density of a mode scales $\propto k_{\perp}^2$ (equation \ref{eq:Wtot}). Therefore, removing the high-$k_{\perp}$ modes forces the low-$k_{\perp}$ modes to grow to larger amplitudes to achieve the same amount of beam angular broadening. The net effect is that LLD allows a larger portion of the beam’s kinetic energy to be channeled into the plasma oscillations.

Figure \ref{fig:delta_theta} tracks the time evolution of the pair-beam’s angular spread at 20 Mpc for several representative Lorentz factors. To be noted from the figure is that, despite the differences in the spectrum, the instability is still saturated by the transverse diffusion, and the beam eventually reaches a comparable steady-state angular spread with or without LLD. We also notice that with LLD (solid curves) beam scattering proceeds slightly later than without LLD (dotted curves). This slight delay is enough to allow the exponential grow of the instability to achieve higher oscillation amplitudes with LLD.

Figure~\ref{fig:delta_theta} demonstrates that at saturation the angular spreads with LLD are only slightly lower than those without. Hence there is a minimal impact on the arrival time delay of the GeV cascade emission. We found that the cascade time delays in the presence of LLD are smaller by only 2\% to 4\% than they are without LLD \citep{Alawashra_2025}, depending on the cascade photon energy.

Including LLD increases by more than a factor ten the beam energy loss due to the instability. This is demonstrated in figure \ref{fig:dUdt} that shows the time evolution of the magnitude of the beam energy loss rate due to the instability without (dotted blue lines) and with (solid blue lines) LLD at selected distances of 20 and 50 Mpc from the source. We calculate the instability loss rates using equation \ref{eq:dUdt} and without accounting for this energy loss feedback on the beam; we only include the scattering feedback of the instability. 

The red lines in figure \ref{fig:dUdt} represent the beam energy input by the pair production rate, which is quasi-constant over the time span considered there. The black lines show energy losses due to inverse Compton (IC) cooling, which we account for in the beam evolution (equation \ref{eq:f}), unlike the instability loss. Over a timescale of a few million years, the instability energy loss in the LLD case asymptotically approaches a steady state, taking about a tenth of the energy dissipation budget. In contrast, without LLD the instability energy loss remains more than two orders of magnitude lower than IC cooling losses.

Note that the IC energy loss rate does not balance the production rate in figure \ref{fig:dUdt}. The reason of this is that the IC cooling time are longer for pairs with lower energy which results in the accumulation of the pairs at lower energies. We neglect those low energy tail pairs every time we initialize the instability feedback calculations as they have negligible impact on the instability energy loss of the high energy pairs relevant for the IC cascade ($\gamma> 10^5$).

Despite this increase in wave energy density with LLD, a quasilinear description of the instability is still justified. At saturation, the fraction of the total wave energy density to the IGM thermal energy density varies from $10^{-5}$ at 5 Mpc from the blazar to $10^{-9}$ at 125 Mpc.

\begin{figure}
    \centering
    \includegraphics[width=\columnwidth]{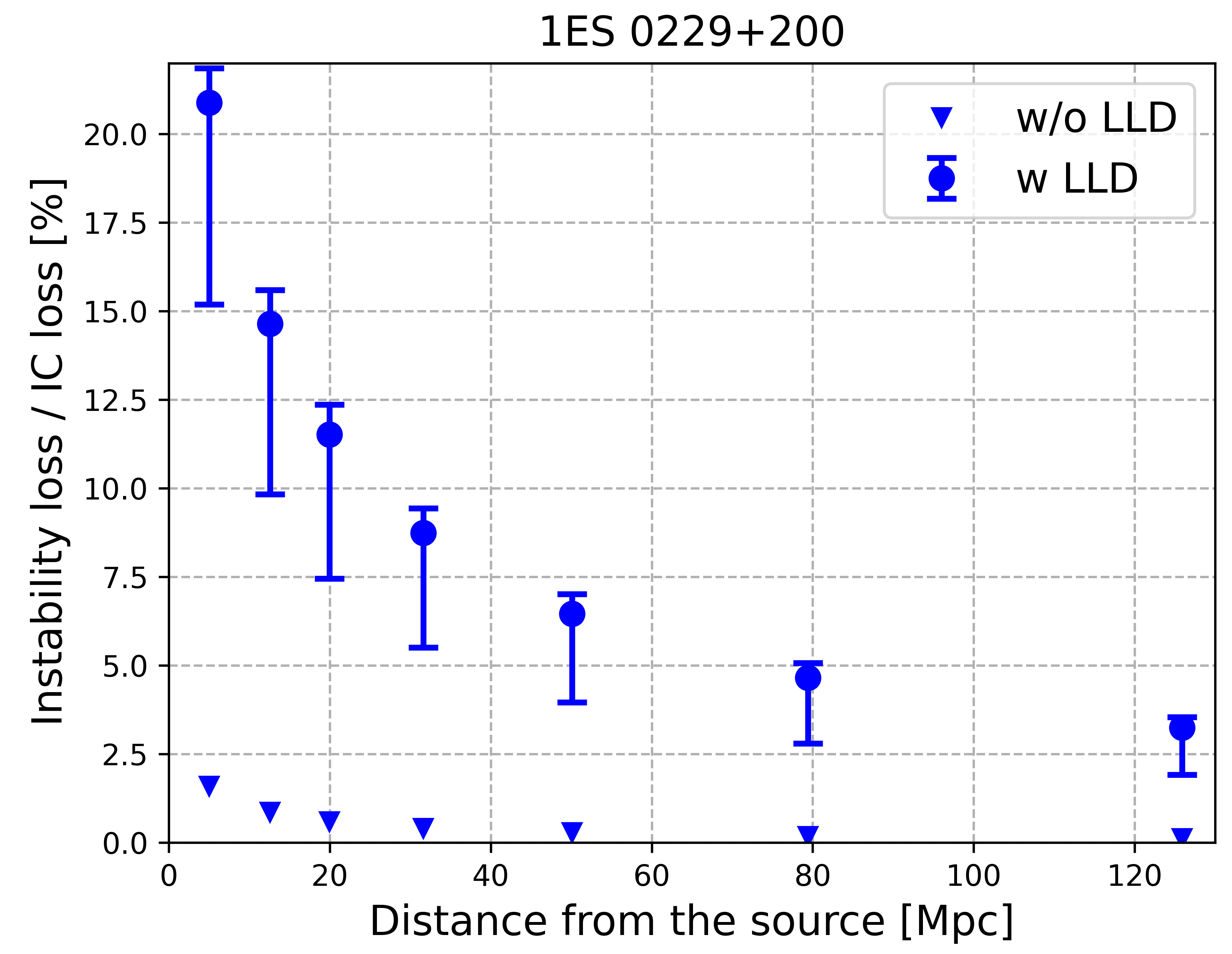}
    \caption{Instability-induced energy-loss fraction relative to inverse Compton (IC) loss as a function of distance from the blazar 1ES 0229+200. The blue circular markers represents the instability loss fraction when LLD is included, showing a significant increase in energy dissipation. The blue triangular markers shows the instability loss without LLD, where the energy loss remains negligible across all distances. The lower and upper error bars of the LLD-inclusion case correspond to the low and high models of the MeV gamma-ray background as described in section \ref{sec:LLD}. We clearly see a substantial enhancement of instability-driven energy dissipation of the pair beams when LLD is considered.}
    \label{fig:W/IC}
\end{figure}

Figure \ref{fig:W/IC} shows the ratio of instability-induced energy loss to that by IC scattering as a function of distance from the blazar 1ES 0229+200. The blue triangular markers represent the energy-loss ratio without LLD, and none of them exceeds the $2-\%$ level. The blue circular markers indicate the energy-loss fraction with LLD. Their significantly higher level is due to the suppression of oblique modes by LLD, allowing quasi-parallel modes to grow more efficiently and extract greater amounts of energy from the beam.

The error bars of the circular markers in figure \ref{fig:W/IC} reflect the uncertainty in the cosmic-ray electron distribution. The markers themselves are for the fiducial MeV gamma-ray background model used in \cite{Yang_2024}. The extent of the error bars corresponds to the "low" model and the "high" model. Both models were introduced in section \ref{sec:LLD} and shown in figure \ref{fig:LDrate}. 

The variation between the low and high models in figure \ref{fig:W/IC}  emphasises how modestly the energy dissipation by the instability depends on the extragalactic cosmic-ray electron background. Reducing the LLD rates by one order of magnitude reduces the instability energy loss rates only by less than a factor of two.

We note that the ratio of the instability loss rates with LLD to those without LLD slowly increases with distance from the blazar. This enhancement is due to a decline of the linear growth rate that reflects the dilution of the pair beam density with the distance squared. The linear growth balances damping (LLD and the collisional damping) at certain perpendicular wave numbers, $k_\perp$, that shift to lower $k_\perp$ as the linear growth decreases. As a result, the wave spectrum shifts to beam-parallel modes that induce less angular scattering and are more efficient in extracting the beam energy.

We show calculations in Figure \ref{fig:W/IC} up to 130 Mpc from the blazar 1ES 0229+200, despite its luminosity distance of 687 Mpc. This is because Fermi-LAT observes cascade emission in the 1–100 GeV range (see Fig.3 in \cite{Aharonian_2023}) that is largely produced by pairs created in the first 130 Mpc from the blazar. For example, 10 GeV cascade photons are produced by IC emission from pairs with $\gamma \simeq 3.4 \times 10^6$. As the CMB is a thick target for the pairs, they must be created with about twice that energy, or $\gamma\simeq 7\times 10^6$, and hence be produced by $7$~TeV gamma rays which have a mean free path of ~80 Mpc. Higher-energy gamma rays are absorbed even closer to the blazar. We also note a rapid decline of the instability-induced energy loss with distance, on account of the $1/R^2$ dilution of the pair beam density. We can extrapolate that the instability does not affect cascade production at distances much beyond 100 Mpc.

We have explored the impact of the IGM density, $n_{e}$, variation the instability efficiency. We found that by reducing the density by a factor of 3.3 the instability to IC loss fraction increases by factors of 1.6, 2.0 and 2.4 for the corresponding distances of 13, 50 and 125 Mpc. This suggest a stronger impact of the IGM density for larger distances. However, the instability efficiency turns weaker with increasing distance due to the declining beam density.

The results presented in this paper were obtained using a intrinsic $\gamma$-ray spectrum and luminosity for a 1ES 0229+200-like blazar. We also explored how variations in the source’s luminosity affect instability energy-loss fraction from the secondary cascade. In particular, we found that at distance 50 Mpc the energy-loss fraction increases from 6.5$\%$ to 10$\%$, if we enhance the luminosity by a factor of $\sqrt{10}\simeq 3$, and to 14.8$\%$ for a luminosity elevated by a factor of ten. Therefore, we conclude that more luminous TeV source (higher pair injection rate) leads to a moderate increase in the energy loss by the electrostatic instability.

It is important to note that in our current model, we track wave growth and beam scattering, but the energy taken up by the plasma waves is not explicitly subtracted from that of the beam. We emphasize that a quantitative assessment of the observable consequences (in terms of the secondary GeV cascade flux from blazars) will require incorporating the beam’s energy loss self-consistently.

\text{      }

\section{Conclusion}\label{sec:con}

We consider the impact of linear Landau damping (LLD) by MeV-scale cosmic-ray electrons in the IGM on the evolution of the blazar-induced pair beam instability. We have performed a numerical calculation of the nonlinear feedback of electrostatic beam-plasma instabilities on the blazar 1ES 0229+200 induced pair beams at different distances in the IGM. We found that LLD dramatically enhances the beam’s energy loss due to the plasma instability, by damping electrostatic modes oblique to the beam.

With the oblique modes suppressed by the LLD, the quasi-parallel modes grow to larger amplitudes and extract a significantly larger fraction of the beam energy. This increases the energy-loss efficiency of the instability by more than an order of magnitude. LLD thus fundamentally changes the instability’s evolution, suggesting that such damping must be accounted for in any comprehensive model of TeV pair-beam propagation in the IGM. For a 1ES 0229+200-like blazar, we found that the fraction of instability-induced energy loss of the beam relative to the energy emitted as IC secondary emission is around 10 - 20 $\%$ for a small distance from the source ($\sim 1 - 20$ Mpc) and less than this for larger distances. 

Predicting the degree to which the GeV cascade flux from 1ES 0229+200 (and similar blazars) would be reduced due to the LLD-enhanced instability requires further steps to the calculations introduced in this paper. Future work will address a momentum-dependence of the instability-induced energy loss of the beam that could modify the distribution function of the beam, and hence the instability growth rate. Constructing predictions from the beam-plasma instability calculations will be useful for cross-checking with observations, as it will tell us whether the instability energy loss is sufficient to explain the missing cascade emission.

\begin{acknowledgments}
This research was
supported by the Munich Institute for Astro-, Particle and
BioPhysics (MIAPbP), which is funded by the Deutsche
Forschungsgemeinschaft (DFG, German Research Foundation)
under Germany's Excellence Strategy – EXC-2094 –
390783311. During the preparation of this work, C.H. was supported by the David \& Lucile Packard Foundation; and by the National Aeronautics and Space Administration (grant 22- ROMAN110011).
\end{acknowledgments}


\bibliography{Main}{}
\bibliographystyle{aasjournalv7}



\end{document}